# Particle simulation code for charge variable dusty plasmas
# B. Farokhi

Institute for studies in theoretical physics and mathematics
P. O. Box 19395-5531, Tehran, Iran
Department of Physics, Bu-Ali Sina University, Hamadan, Iran

Charges on the dust grains immersed in a plasma vary in time. We follow the hydrodynamic approach to obtain dispersion relations for some low frequency electrostatic modes, taking into account charge fluctuation on the dust. We have used the description of the one dimensional fully electromagnetic particle code. There exists one spatial dimension ($x$) and three velocity components $(v_x, v_y, v_z)$. This is usually referred to as a $1\frac{2}{2}$ model. Our analysis shows that the presence of the dust grains causes different effects in different frequency regimes.

## Computational model

This is the description of the one dimensional fully electromagnetic particle code. The descriptions are intended to clarify largly the code write-up to better follow its algorithm in a concise manner.

There exists one spatial dimension ($x$) and three velocity components $(v_x, v_y, v_z)$. This is usually referred to as a $1\frac{2}{2}$ model (e.g. $x$ is considered a full dimension since it retains both spatial and velocity components, while $y$ and $z$ are considered a half dimension each). Electrons and ions though are followed in time using the self consistent electromagnetic fields in the simulations.

The magnetic field **B** has only two components (transverse), $B_y$ and $B_z$, i.e., the transverse components. The electric field though has three components, one longitudinal ($E_x$), and two transverse ($E_y$, and $E_z$).

Waves can only propagate along the $x$ direction. For the longitudinal modes the electric field disturbance ($E_x$) is the wave generator, while to the transverse modes the transverse ($E_y$, $E_z$) as well as the transverse magnetic fields ($B_y$ and $B_z$) are the wave generators. The longitudinal and the transverse fields are advanced separately using Poissons equations and the coupled Amperes and Faradays laws.

## Analytic treatment

The evolution of the trajectories of the electrons in their self consistent electromagnetic fields can be approximated to zeroth order by the linearized nonrelativistic fluid equations

$$-n_0 e \frac{d\mathbf{v}}{dt} = \frac{n_0 e^2}{m}\mathbf{E} = \frac{d\mathbf{j}}{dt} = \frac{w_{pe}^2}{4\pi}\mathbf{E} \qquad (1)$$

where $m$, $n_0$, **v** and $e$ represent the electron mass, background density, flow velocity and the electric charge respectively; $w_{pe}$ is the electron plasma frequency.

The current and the longitudinal part of the electric field determined from the following

$$\mathbf{j} = (-n_e \mathbf{v}_e + n_i \mathbf{v}_i)e \qquad (2)$$
$$\nabla \cdot \mathbf{E} = -4\pi e(n_e - n_i) \qquad (3)$$

where, $n_e$ ($n_i$) represents the electron (ion) density. The transverse parts of the electric and magnetic fields are determined from the coupled Faraday's and Ampere's laws as follows

$$\nabla \times \mathbf{E} = -\frac{1}{c}\frac{\partial \mathbf{B}}{\partial t} \qquad (4)$$

$$\nabla \times \mathbf{B} = \frac{1}{c}\frac{\partial \mathbf{E}}{\partial t} + \frac{4\pi}{c}\mathbf{j} \qquad (5)$$

Noting that in the one dimensional problem only $\partial/\partial x$ is nonzero, then from these the following pair of coupled equations results

$$\frac{\partial E_y}{\partial x} = -\frac{1}{c}\frac{\partial B_z}{\partial t} \qquad (6)$$

$$\frac{\partial B_z}{\partial x} = -\frac{1}{c}\frac{\partial E_y}{\partial t} - \frac{4\pi}{c} j_y \qquad (7)$$

Eliminating $B_z$ from the Eqs. 6 and 7 will lead to the dispersion relation of the electromagnetic waves as follows

$$\frac{\partial^2 E_y}{\partial x^2} = -\frac{1}{c}\frac{\partial^2 B_z}{\partial x \partial t} = \frac{1}{c^2}\frac{\partial^2 E_y}{\partial t^2} + \frac{w_{pe}^2}{c^2} E_y \qquad (8)$$

Assuming simple plane wave solution in the Eq. 8 for $E_y$ (e.g. $E_y \propto \exp[i(kx - wt)]$) will give rise to the electromagnetic dispersion relation of the waves propagating in the plasma medium as follows

$$w^2 = w_{pe}^2 + k_z^2 c^2 \qquad (9)$$

It can be seen that repeating the above procedure for other transverse components of the electric or magnetic field will result in the same dispersion relation as that derived in 9.

As regards the longitudinal waves, performing Fourier transform of the momentum and Poisson's equation and using the continuity equation for the electrons one can simply eliminate $E_x$ and show the following dispersion relation

$$w = w_{pe} \qquad (10)$$

Of course had we used first rather than zeroth order approximation to the momentum equation, we would simply obtain the Langmuir wave dispersion relation as follows

$$w^2 = w_{pe}^2 + k_z^2 v_{th}^2 \qquad (11)$$

Using the ion motion cause ion acoustic wave.

In what follows, therefore we shall test the model via the longitudinal dispersion relation by power analyzing the $E_x$, and the transverse dispersion relation by power analyzing any of the transverse components $E_y, E_z, B_y, B_z$.

### Numerical algorithm

Our equations will be divided into two main classes, the field equations and the particles equations. The model is normalized such that the electron charge to mass ratio $(e/m)$ is unity. Furthermore, imposing $4\pi n_0 e = 1$ will ensure $w_{pe} = 1$, which is used as the fundamental frequency of the plasma oscillations and enters both the transverse as well as the longitudinal dispersion relation. This has the added advantage that the step can be represented in terms of $1/w_{pe}$.

Spatial differences of the field equations are all done in the $k$ space. As such then the field equations temporal dependence are the following

$$E_x = -i\frac{n}{n_0}\frac{\exp(-k_x^2 a_x^2)}{k_x} \qquad (12)$$

$$\frac{\partial E_y}{\partial t} = \frac{n}{n_0}\exp(-k_x^2 a_x^2) j_y - ik_x c B_z \qquad (13)$$

$$\frac{\partial E_z}{\partial t} = \frac{n}{n_0}\exp(-k_x^2 a_x^2) j_z - ik_x c B_y \qquad (14)$$

$$\frac{\partial B_y}{\partial t} = ik_x c E_z \qquad (15)$$

$$\frac{\partial B_z}{\partial t} = ik_x c E_y \qquad (16)$$

Note that the only physical constant which remains intact of the normalization is the speed of light $c$; all the spatial derivatives with respect to x are replaced by $ik_x$; and the factor of $\exp(-k_x^2 a_x^2)$ ($a_x$ represents the filtering scale) is introduced in front of all the interpolated quantities in $k$ space to suppress high $k$ (short wavelength) noise resulting from the linear interpolations of the charge and currents on the mesh and simply acts as a smoothing factor. Another way of understanding this is, this quantity is simply the particle finite size shape factor. Also with the normalization chosen, the current and the velocity interpolated to the mesh from the particle distributions are the same.

As regards the particle pusher, the three velocity difference equations in time centered form can be written as follows

$$\frac{v_x^{n+1} - v_x^n}{\Delta t} = E_x^{n+1/2} + \frac{v_y^{n+1} + v_y^n}{2c} B_z^{n+1/2} - \frac{v_z^{n+1} + v_z^n}{2c} B_y^{n+1/2} \qquad (17)$$

$$\frac{v_y^{n+1} - v_y^n}{\Delta t} = E_y^{n+1/2} - \frac{v_x^{n+1} + v_x^n}{2c} B_z^{n+1/2} + \frac{v_z^{n+1} + v_z^n}{2c} B_x^{n+1/2} \qquad (18)$$

$$\frac{v_z^{n+1} - v_z^n}{\Delta t} = E_z^{n+1/2} + \frac{v_x^{n+1} + v_x^n}{2c} B_y^{n+1/2} - \frac{v_y^{n+1} + v_y^n}{2c} B_x^{n+1/2} \qquad (19)$$

Here the superscripts represent temporal difference indices, and clearly the objective is to go from the step $n$ to $n+1$, with quantities at $n+1/2$ as the middle step quantities used to time center the difference equations and therefore make the time advancement second order accurate. This set can be solved exactly for all the $n+1$ dependent $v$'s, by taking all the $n+1$ dependent $v$'s to the left and all the $n$ dependent $v$'s to the right, and use matrix inversion. This is how the model pushes particles in time.

Finally the last but not the least item in this section is the interpolation from the particles to the mesh and from the mesh to the particles. In this version of the code the so called nearest grid point interpolation which is a zeroth order interpolation is implemented; i.e., the charge and currents of a particle are completely given to the grid point closest to it. This procedure is indeed a very noisy and dissipative one, but in order to make it somewhat smoother, in this code, there are contributions made to the grid points on the left and right of the nearest grid point; these contributions are of equal magnitude and opposite sign, with the positive contribution going into the grid point closer to the particle and the negative contribution to the grid point further apart from the particle. Particles on the other hand receive electric and magnetic fields from the nearest grid points by the same interpolation from their nearest grid point. The symmetry of interpolation between the particles and the grid points insures momentum conservation and zero (to round off) self-forces.

**Conclusions**
We have shown that the coupling of charge relaxation of the dust grains with Langmuir waves leads to decay of the latter. That is, there can be energy transfer between the plasma waves and the dust charging process. Similar decay are expected to occur for high frequency waves. It is not clear how the present decay saturated. Over longer times, we have the short growing stage, the usual quasi linear and nonlinear mechanisms fr decay saturation, here there is also the possibility of enhanced shielding of the dust grains.

We should emphasize that result such as the present one are strongly dependent on the model of the dusty plasma. In particular case, when the total charge is conserved, the total number of electrons and ions conserved. In the other model with infinite plasma, the number of electrons and ions is conserved, so the charge is not conserved.

Figure 1 shows the dust charge versus time. The dust charge is negative but in the figure absolute value of charge is expected. The vertical axis measured by $10^4 e$ ($e$ is the charge of electron) and the horizontal axis measured by 50 time step. Figure 2 shows the power spectrum of the $E_z(k_x = 20, k_y = 0)$ for the time step 400 (solid curves) and for the time step 600 (dotted curves). By normalization of frequency, $w \cong w_p = \pm 1.0$ are chosen.

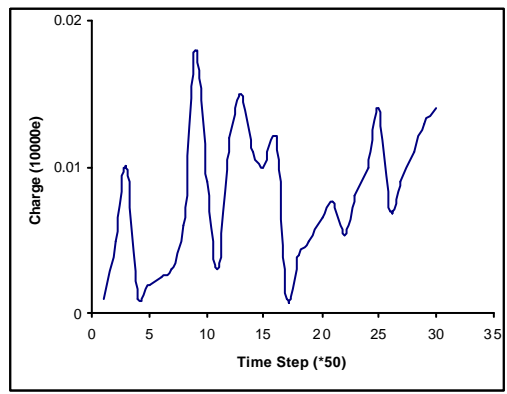

Figure 1: Charge of dust versus time step

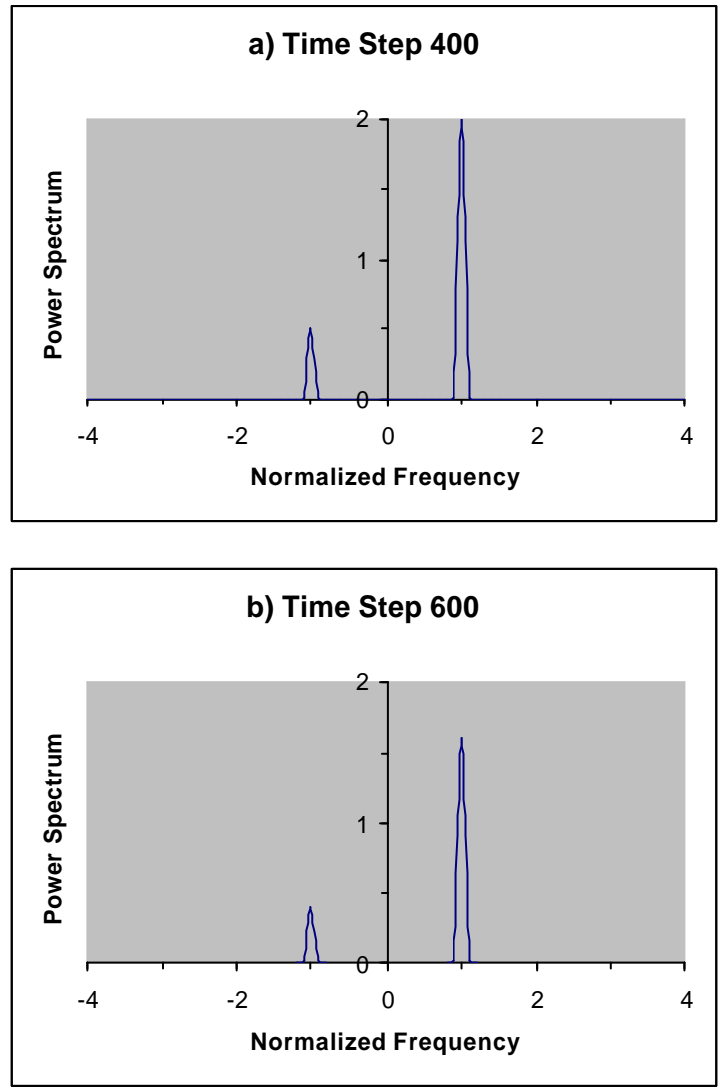

Figure 2: Power spectrum versus frequency
Time step 400
Time step 600